\documentclass[twocolumn,aps,pre,showpacs]{revtex4-1}
\usepackage{amsmath,amssymb}
 \usepackage{graphicx}
\usepackage{hyperref}
\usepackage{color}

\newcommand{\light}{\mathcal A}
\newcommand{\shade}{\mathcal P}
\newcommand{\rhol}{\rho_{\scriptscriptstyle\mathcal A}}
\newcommand{\rhos}{\rho_{\scriptscriptstyle\mathcal P}}

\newcommand{\Vs}{V_{\scriptscriptstyle\mathcal P}}
\newcommand{\Vt}{V}

\newcommand{\Ns}{N_{\scriptscriptstyle\mathcal P}}
\newcommand{\Nt}{N}

\newcommand{\Pe}{\mathrm{Pe}}
\newcommand{\X}{\mathcal{\ell}}
\newcommand{\eg}{e.g.\ }
\newcommand{\egg}{e.g.}

\newcommand{\pt}{\tau_p}

\renewcommand{\pl}{\ell_p}
\newcommand{\pll}{\ell_{p,\scriptscriptstyle\mathcal A}}

\newcommand{\Dr}{D_\mathrm{r}}
\newcommand{\Deff}{D_\mathrm{eff}}
\newcommand{\vsp}{v_\mathrm{sp}}
\newcommand{\gammar}{\gamma_\mathrm{r}}
\begin{document}
\title{Trapping of interacting propelled colloidal particles in
  inhomogeneous media}
\author{Martin P. Magiera}
\author{Lothar Brendel}
\affiliation{Faculty of Physics and CENIDE, University of Duisburg-Essen,
D-47048 Duisburg, Germany
}

\date{\today}
\pacs{82.70.Dd,05.40.Jc, 05.70.Ln, 47.57.-s}
\begin{abstract}
  A trapping mechanism for propelled colloidal particles based on an
  inhomogeneous drive is presented and studied by means of computer
  simulations. In experiments this method can be realized using
  photophoretic Janus particles driven by a light source, which shines
  through a shading mask and leads to an accumulation of the particles
  in the passive part. An equation for an accumulation parameter is
  derived using the effective inhomogeneous diffusion constant
  generated by the inhomogeneous drive. The impact of particle
  interaction on the trapping mechanism is studied, as well as the
  interplay between passivity-induced trapping and the emergent
  self-clustering of systems containing a high density of active
  particles.  The combination of both effects makes the clusters more
  controllable for applications.
\end{abstract}

\maketitle

\section{Introduction}
In recent years, various approaches have been pursued to construct
propelled colloidal particles, including man-made active swimmers at
the micron scale.  Such objects undergo Brownian motion driven by
equilibrium thermal fluctuations, but are also driven by a propulsion
mechanism out of equilibrium, resulting in an interplay between
equilibrium thermal motion and non-equilibrium drive \cite{Hanggi2009,
  Romanczuk2012}. Besides the construction of artificial flagella
\cite{Dreyfus2005} which mimic natural microswimmers like bacteria,
algae or sperm or the use of molecular machines \cite{Mavroidis2004},
novel nano motors have been proposed to drive passive colloids.  An
elegant way to bring passive colloids into motion is the generation of
chemical, electrostatic or thermal field gradients \cite{Anderson1989,
  Piazza2008} leading to a net propulsion via chemophoresis,
electrophoresis or thermophoresis. Introducing an inhomogeneous
coating of the colloidal particle leads to a self-generated slip
velocity pattern at the colloidal surface \cite{Julicher2009},
inducing a net propulsion of particles via self-thermophoresis \cite{
  Jiang2010, Baraban2013} or self-diffusiophoresis
\cite{Golestanian2005, Golestanian2007, Howse2007a,
  Volpe2011,Buttinoni2012}. For the case of thermophoresis, the
inhomogeneous absorption of light causes inhomogeneous heating of the
particle, which again leads to a temperature gradient in the medium
and to propulsion, called photophoresis. While most diffusiophoretic
swimmers based on catalytic reactions require a fuel in the liquid
(typically hydrogen peroxide; a fuel free exception is \eg a swimmer in a
binary mixture near the critical demixing temperature used in
Refs.~\cite{Volpe2011,Buttinoni2012}), swimmers based on photophoresis
are fuel free, which makes them suitable for in-vivo applications.

The control of objects in liquids is of fundamental interest. Instead
of passive colloids, propelled colloidal particles move autonomously
in the liquid where they can fulfill a task, \egg, a catalytic
reaction taking place at the swimmer surface in a lab-on-the-chip
device or in nature to cure an ecosystem (bioremediation). External
control like steering, trapping or sorting of active matter can
facilitate the swimmers' remove from the system after the task is
fulfilled. One realization is the introduction of a geometrical trap
as proposed in Ref.~\cite{Kaiser2012} (or Ref.~\cite{Mijalkov2013} for
chiral swimmers) which collects all active particles after the task is
fulfilled. In recent experiments, passive nano-objects have been
trapped thermophoretically in an inhomogeneous temperature field
generated by gold patterns illuminated by a focused laser beam
\cite{Braun2013}. In another work, active photophoretic swimmers have
been trapped using a complex feedback control mechanism
\cite{Cichos2013}. Therefore, the orientation of individual swimmer
has been recorded. According to the orientation, swimmers are
accelerated when they aim at a predefined target.

\begin{figure}[b]
\centering
\includegraphics[width=\columnwidth]{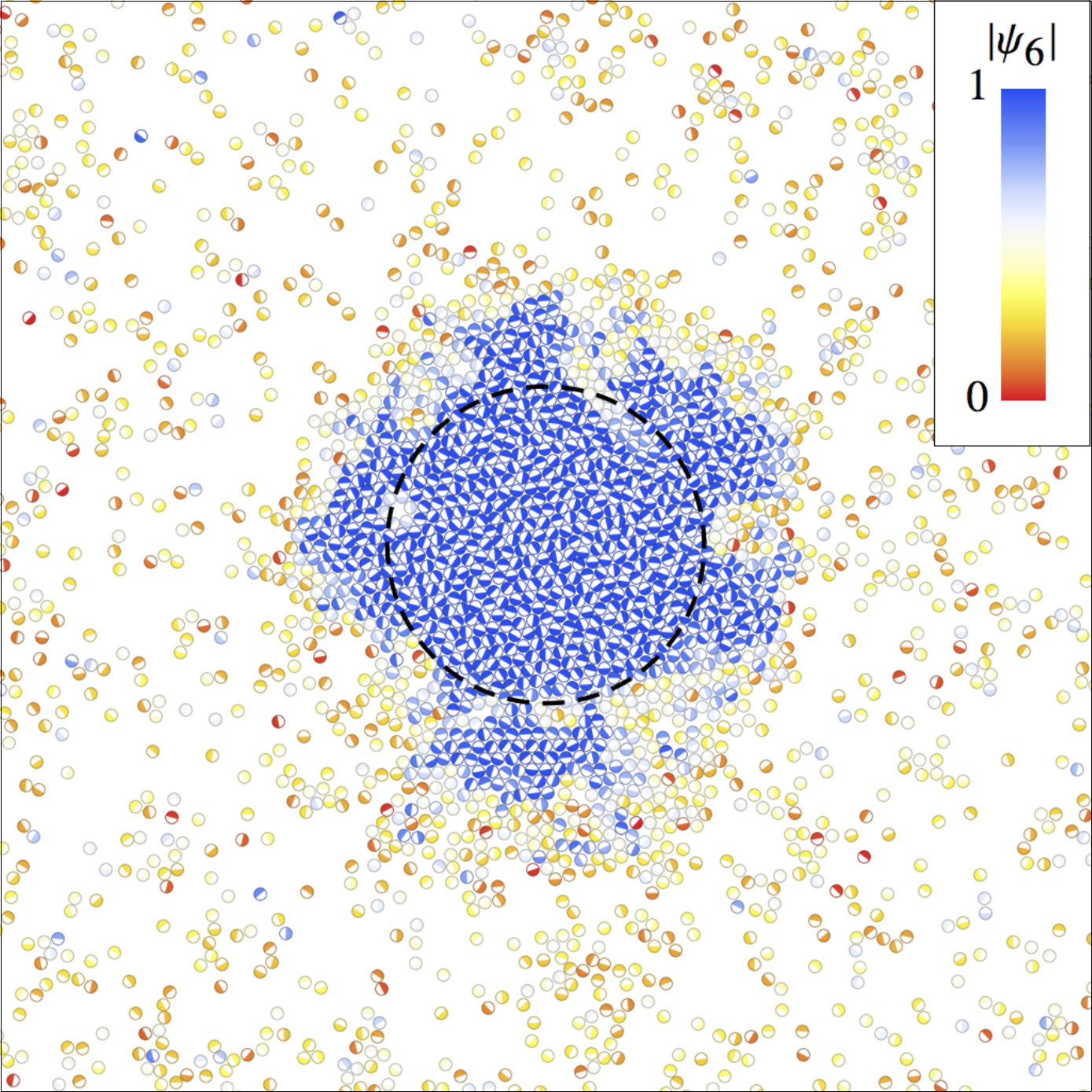}
\caption{\label{fig:sketch} Snapshot of a simulation for a system of
  interacting active particles (at $\Pe {=} 250$ and a particle
  density of $\phi_p {=} 0.25$ defined in Eq~\eqref{eq:phi_p}), which
  are passivated in a spherical region and accumulate there. The color
  coding shows the bond order parameter, which is introduced in
  Eq.~\eqref{eq:Psi6} and quantifies local crystallization of
  particles.}
\end{figure}
Using theory and computer simulations, we study the trapping of
particles by only manipulating the propulsion mechanism (or in a
biological way by manipulating their tumbling rate) without any
feedback control mechanism. Whenever swimming objects interact with
each other, their dynamics is influenced by the interaction. Recently,
it has been shown that system which interact only via repulsive
interactions can accumulate driven by activity when driving and
density are large enough. This \emph{athermal phase separation}
\cite{Fily2012, Redner2013, Redner2013b, Buttinoni2013,Stenhammar2014}
is related to our study as it is based on inhomogeneities of, e.g.,
the diffusion constant in regions, where spontaneously a higher
particle density exists. However, in solely active systems, these
inhomogeneities are generated dynamically and cannot be controlled
externally. Turning off the propulsion in a part of the system leads
to an accumulation of particles, a fact which is known
\cite{Schnitzer1990, Schnitzer1993, Cates2012} and observed in
experiments \cite{Kuemmel2014} but has not been addressed explicitly
in an own publication to the knowledge of the authors. We show that
the local \emph{marking} of the target region can be used to induce
the emergent athermal phase separation, which enhances the effect of
accumulation.

\section{Model}
We describe Janus particles suspended in a viscous liquid using
Brownian Dynamics, which is Langevin dynamics in the overdamped limit. This limit is essential in the low Reynolds number
limit \cite{Purcell1977, Berg1993}. It is governed by the equations
for position and orientation of particle $i$:
\begin{subequations}
\begin{align}
\dot{\mathbf r}_i &= \sqrt{2 D} \boldsymbol{\eta}_i(t) + \mathbf F_i/\gamma \\
\boldsymbol {\omega}_i & = \sqrt{2 \Dr} \boldsymbol{\eta}_{i}^\mathrm{r}(t) + \mathbf T_i/\gammar  \\
\dot{\hat{\mathbf o}}_i &= \boldsymbol{\omega}_i \times \hat{\mathbf o}_i,
\end{align}
\end{subequations}
with $\mathbf F_i$ and $\mathbf T_i$ representing all forces and
torques acting at particle $i$, the latter being neglected in this work. The vectors $\boldsymbol{\eta}_i(t)$ and
$\boldsymbol{\eta}_i^\mathrm{r}(t)$ contain uncorrelated, normally distributed
noise with zero mean and second moment $1$
\cite{Uhlenbeck1930}, which is motivated by a strict separation of
the timescales of the particle friction due to particle
liquid collisions and particle motion. The spatial diffusion coefficient $D$
depends on temperature and Stokes' damping $\gamma$ via $D {=}
\gamma/(k_\mathrm{B} T)$. For the rotational diffusion coefficient
(spherical particles) the
particle diameter $\sigma$ enters via $\Dr = 3 D/\sigma^2 =
\gammar/(k_\mathrm{B} T)$ if we
assume no slip between a particles' surface and the liquid
\cite{Einstein1906}. The equations of motion can be solved using
Euler's or higher order methods.

Having introduced a model for spherical particles suspended in a
solvent, we now introduce a propulsion via a self-propulsion velocity
$\vsp$ in terms of an effective force, as well as interactions via a pair
potential $\Phi$,
\begin{equation}
  \mathbf F_i = \gamma \; \vsp(\mathbf r_i)  \hat{\mathbf o}_i -
  \sum_j\nabla_i \Phi(\mathbf r_i-\mathbf r_j).
\label{eq:force}%
\end{equation}
Thus, the self-propulsion velocity points into the direction of a particle's
orientation $\hat {\mathbf o}_i$ and depends on the particle's position
$\mathbf r_i$. The two motion mechanisms, passive thermal diffusion
and active motion due to propulsion, are compared in the dimensionless
P\'eclet number
\begin{equation}
\Pe = \frac{\sigma \vsp}{D},
\end{equation}
which is of fundamental importance in this work.

As a pair potential we use the truncated Lennard-Jones potential \cite{Weeks1971},
\begin{align}
 \Phi(\mathbf r) 
&= \left \{ \begin{array}{ll}  4 \epsilon \left ( \left (\frac{\sigma}{r} \right )^{12} {-} \left (\frac{\sigma}{r} \right )^{6} \right ) {+} \epsilon & r \le 2^{1/6} \sigma  \\
0 & r > 2^{1/6} \sigma \end{array} \right .,
\end{align}
which contains only the repulsive core of the common Lennard-Jones
potential. For the interaction energy we choose either $\epsilon {=} 100
k_\mathrm{B}T$ or zero for the case of no interaction. In regions of
large particle densities, particles can crystallize with a lattice
constant starting of $2^{1/6} \sigma$ at intermediate density and
slightly decreasing with rising density. Note that when disregarding
rotational Brownian motion, actually the particles do not have a
physical \emph{extension}, but only define a length caused by
interaction.

\section{System geometry}
We study a system with two regions: In the active one, $\light$,
objects are propelled ($\vsp(\mathbf r) = \vsp^0 > 0$) while in the
other, $\shade$, only thermal diffusion takes place since
$\vsp(\mathbf r) = 0$. Because this is the only inhomogeneity type of
$\vsp(\mathbf r)$ considered and for the sake of brevity, we denote in
the following the constant $\vsp^0$ by $\vsp$ as well. The
corresponding propulsion of particles in the illuminated region
prevents the system to reach equilibrium: In the non-equilibrium
steady state, particles accumulate in the passive region, as shown in
Fig.~\ref{fig:sketch}.

To quantify the accumulation, it is convenient to evaluate the
difference between the densities in the passive region $\shade$ and the
active region $\light$, which are defined by
\begin{equation}
  \rhos = \frac{\Ns}{\Vs}
  \quad\text{and}\quad
  \rhol = \frac{\Nt-\Ns}{\Vt-\Vs},
\end{equation}
where $\Nt$ and $\Vt$ are the total number of particles and total
volume of the system, respectively.
For the case that all particles are captured in region $\shade$, the
density difference becomes $\Nt/\Vs$. We use
this maximum value to normalize the density difference, and define the
accumulation parameter
\begin{equation}
\Delta =\frac{\rhos-\rhol}{\Nt/\Vs} =
\frac{\frac{\Ns}{\Nt}-\frac{\Vs}{\Vt}}{1-\frac{\Vs}{\Vt}}.
\label{eq:opam}
\end{equation}
Now, this parameter is unity for the case that all particles are
trapped in the passivation volume of relative size $\phi {=} \Vs/\Vt$
while it is zero for a homogeneous swimmer distribution (to be
expected in the limit $T {\to} \infty$). In a finite system, it can
even become negative due to fluctuations. Note as well that due to the
normalization, $\Delta$ vanishes in the thermodynamic limit
($V{\to}\infty$, $\Vs$ fixed), which we do not consider here, though.

The
accumulation parameter can be also written in terms of the excess
density $\alpha$ defined by $\rhos = \rhol(1+\alpha)$ or
\begin{align}
\label{eq:def_alpha}
\alpha &= \frac{\rhos}{\rhol}-1 \\
\label{eq:Delta}
\Delta &= \frac{1}{1+\frac{1}{\alpha \phi}}.
\end{align}
In the following, we will derive an expression for the accumulation
parameter $\Delta$. We assume that the particle density in the active
region $\light$ is homogenous. The homogenization is driven by thermal
diffusion $D$ (as it is in the passive region $\shade$) as well as the
combination of random rotation $\Dr$ and driving $\vsp$.

\section{Accumulation}
\subsection{Low particle density: No interaction}
Let us first assume the case of low particle density, where
particle-particle interaction can be neglected. The relevant
physics take place at the boundary between $\shade$ and
$\light$. Particle flow due to random motion is described by Fick's
first law of diffusion which usually leads to a homogeneous density
profile in the steady state. Here, however, the underlying system
shows an inhomogeneity: In $\light$ the particles are driven and are
subjected to random rotation, which introduces a \emph{persistence
  length} $\pll$ after which the previous particle orientation is
forgotten. On larger scales this corresponds to Brownian motion with
steps of size $\pll$. In $\shade$ the step size, stemming solely from
$D$ is smaller. Hence, we get an inhomogeneous effective diffusion
coefficient $\Deff$ as derived below.

Taking this inhomogeneity due to a gradient in step size into account,
we arrive at an extended Fick's law of diffusion \cite{Schnitzer1990,
  Schnitzer1993},
\begin{equation}
J = -\Deff \nabla \rho - \frac{\rho}{2} \nabla \Deff,
\label{eq:extended_fick}
\end{equation}
to be used  in the description of a particle flow from one region to
the other. In the steady state the total flux vanishes, and
the particle injection through the second term is completely balanced
by the usual diffusion current. The solution of this flux equation is
\begin{equation} 
\rho \sqrt{\Deff} = \mathrm{const.} \;\;\;\; \Rightarrow \;\;\;\;
\frac{\rhos}{\rhol} = \sqrt{\frac{D_\mathcal{A}}{D_\mathcal{P}}} = \alpha+1,
\label{eq:inh_D}
\end{equation}
where $\rho_i$ and $D_i$ are homogeneous particle densities and
effective diffusion constants in the corresponding
regions.

In $\shade$ the effective diffusion equals the thermal $D$. We now
restrict ourselves to a two-dimensional system of self-propelled
particles \footnote{Note that Stokes' friction coefficient $\gamma$
  exists strictly only in three dimensions. Actually, calculating the
  friction coefficient of a disk moving in two dimensions would lead
  to the Stokes paradoxon \cite{Stokes, lamb1945hydrodynamics}.} with
a self-propulsion velocity $\vsp$ and persistence time $\pt {=} 2/\Dr$
\footnote{Sometimes in literature, the persistence time is introduced
  using $1/\Dr$. However, a derivation from the Langevin equation
  (e.g. conducted in Ref.~\cite{TenHagen2011}) leads to the transition
  from ballistic to diffusive behavior in Eq.~\eqref{eq:MSD} at
  $2/\Dr$, which makes this definition of $\pt$ more reasonable.},
where the mean square displacement is given by (cf.\ also
\cite{Howse2007a})
\begin{equation}
\langle (\mathbf r(t) - \mathbf r(0))^2 \rangle = 4 D t + \frac{(\vsp
  \pt)^2}{2} \left ( \frac{2 t}{\pt} + e^{-\frac{2 t}{\pt}} - 1 \right ).
\label{eq:MSD}
\end{equation}
This includes the known limits of ballistic motion at short time scales and
diffusive motion with an effective diffusion constant at long time scales.  
Thus, we get in the active region $\light$ the
effective diffusion constant
\begin{equation}
D_\light = D +\frac{\pt \vsp^2}{4} = D \left ( 1+\frac{\Pe^2}{6}
\right )
\end{equation}
leading to 
\begin{equation}
\alpha = \sqrt{1+\frac{\Pe^2}{6}}-1.
\label{eq:sol_od}
\end{equation}

To confirm this result we performed several computer simulations of a
two-dimensional system of 100 non-interacting mono-sized particles in
a simulation cell with a region $\shade$ of $\phi {=} \pi/25$. The
stochastic equations of motion have been solved using Euler's method. 
\begin{figure}[bt]
\includegraphics[width=\columnwidth]{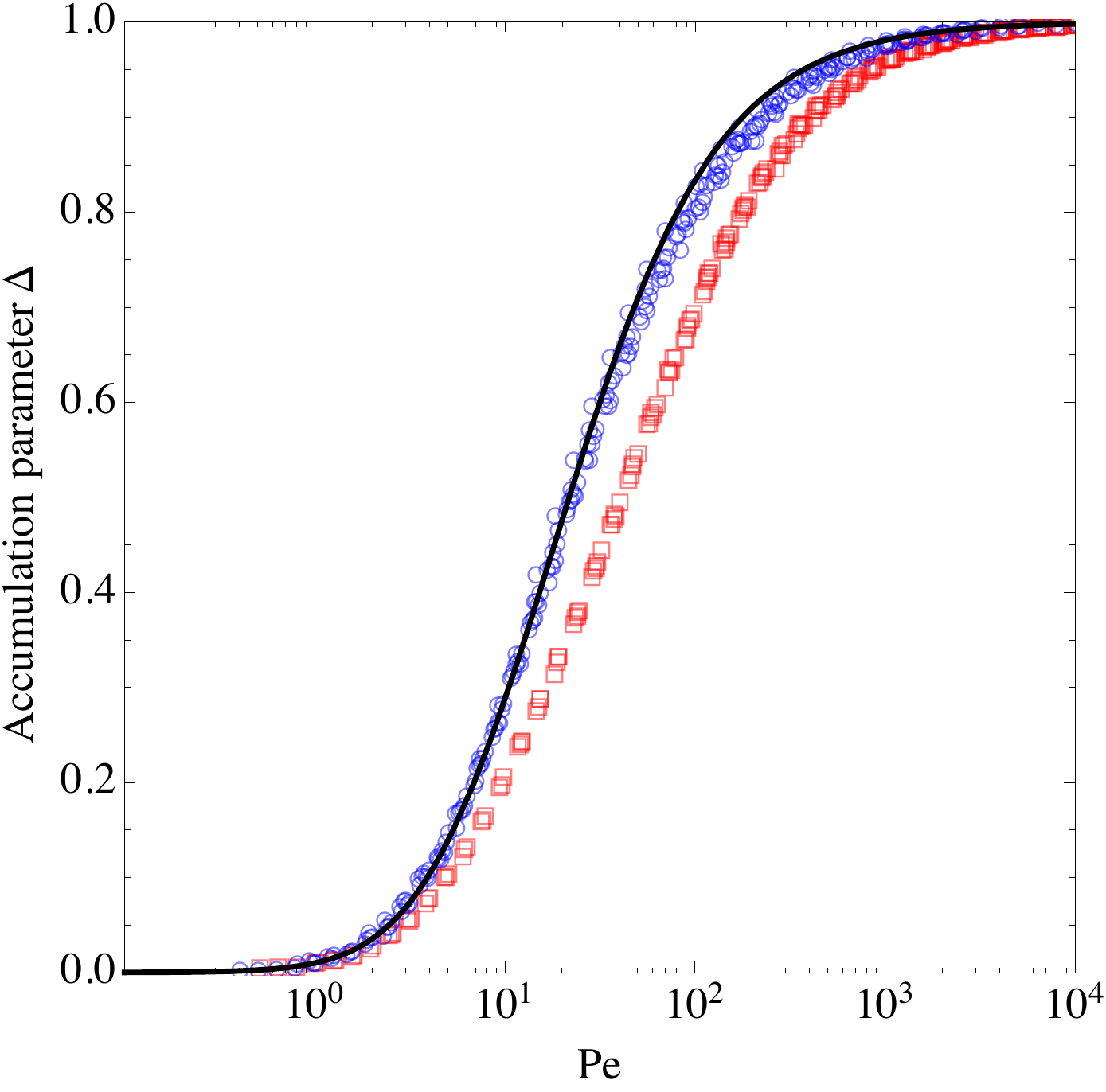}
\caption{\label{fig:overdamped}
  Simulation results for a system of not interacting propelled
  colloidal particles with different $\Pe$. The blue circles
  correspond to systems with $\pl/\X \approx 1/32$, the red squares
  to systems with $\pl/\X \approx 1/4$ (cf.\ Eq.~\eqref{eq:pl}). The
  curve is the result Eq.~\eqref{eq:sol_od} plugged into
  Eq.~\eqref{eq:Delta}.}
\end{figure}

In Fig.~\ref{fig:overdamped} we observe agreement between the
theoretical estimate and results of computer simulations. However, the
theory with solely diffusion coefficients works well on length scales
large compared to the persistence length $\pl$. The introduction of a
finite length by virtue of the width $\ell$ of $\light$ can become
important when $\ell$ is not large compared to the persistence length
\begin{equation}
  \pl=2\sqrt{\Deff\pt}.
  \label{eq:pl}%
\end{equation}
The reason is that the drop of the density between the passive and the
active region from $\rhos$ to $\rhol$ is not sharp, but takes place on the
length scale $\pl$ in the active region: In the active region a
boundary layer is created which scales with $\pl$. No matter where the
boundary layer is formed, any contribution of swimmers to the boundary
layer reduces accumulation. This effect is negligibly small for huge
system sizes ($\ell {\gg} \pl$), but leads to a severe drop of
$\Delta$ in the other limit, where the density $\rhol$ as defined in
Eq.~\eqref{eq:inh_D} is no longer reached.

The here described accumulation in passive regions is related to the
athermal phase separation of interacting particles \cite{Fily2012,
  Redner2013, Redner2013b, Buttinoni2013}, where the collision of
particles leads to a reduced effective diffusion coefficient depending
on the particle density, which again results in the emergent creation
of clusters above a certain packing fraction (cf.\ below). However,
the here described phenomenon occurs at any density and the region of
accumulation is controlled; any spontaneously created clusters form
and dissolve dynamically. The interplay between self-clustering and
trapping is discussed in the next paragraph.

\subsection{Increased particle density}
While in the preceding section the interaction between particles was
neglected, we now consider particles with a hard core repulsion as
introduced in Eq.~\eqref{eq:force}. The interaction introduces --
additionally to the persistence length -- a second length-scale given
by the mean particle distance which depends on the particle
density. Because we use a strong repulsion ($\epsilon =
100k_\mathrm{B}T$) and the mean particle distance in closed packed
  system depends on the particle density, we choose as a length scale
  the truncation length $2^{1/6}\sigma$, and define a particle volume
  fraction
\begin{equation}
\phi_p = \frac{N \pi \sigma^2 2^{1/3}}{V}.
\label{eq:phi_p}%
\end{equation}
Analogous, we may define a volume fraction in the passive region,
$\phi_{p,\mathcal P}$, and one in the active region $\phi_{p,\mathcal A}$.

We may now deduce a maximal accumulation of particles using the finite
\emph{capacity} of the passive region -- once the passive region is
filled, no more active particles can enter.  At least when the local
volume fraction in the passive region exceeds the hexagonal closed
packing value with lattice constant $2^{1/6}\sigma$, the accumulation
is \emph{reduced} compared to the noninteracting case. This is the
case for particle volume fractions exceeding
\begin{equation}
\phi_p^\mathrm{max} = \phi \frac{\pi}{2 \sqrt{3}}.
\label{eq:phi_p_max}
\end{equation}
\begin{figure}[bt]
\includegraphics[width=\columnwidth]{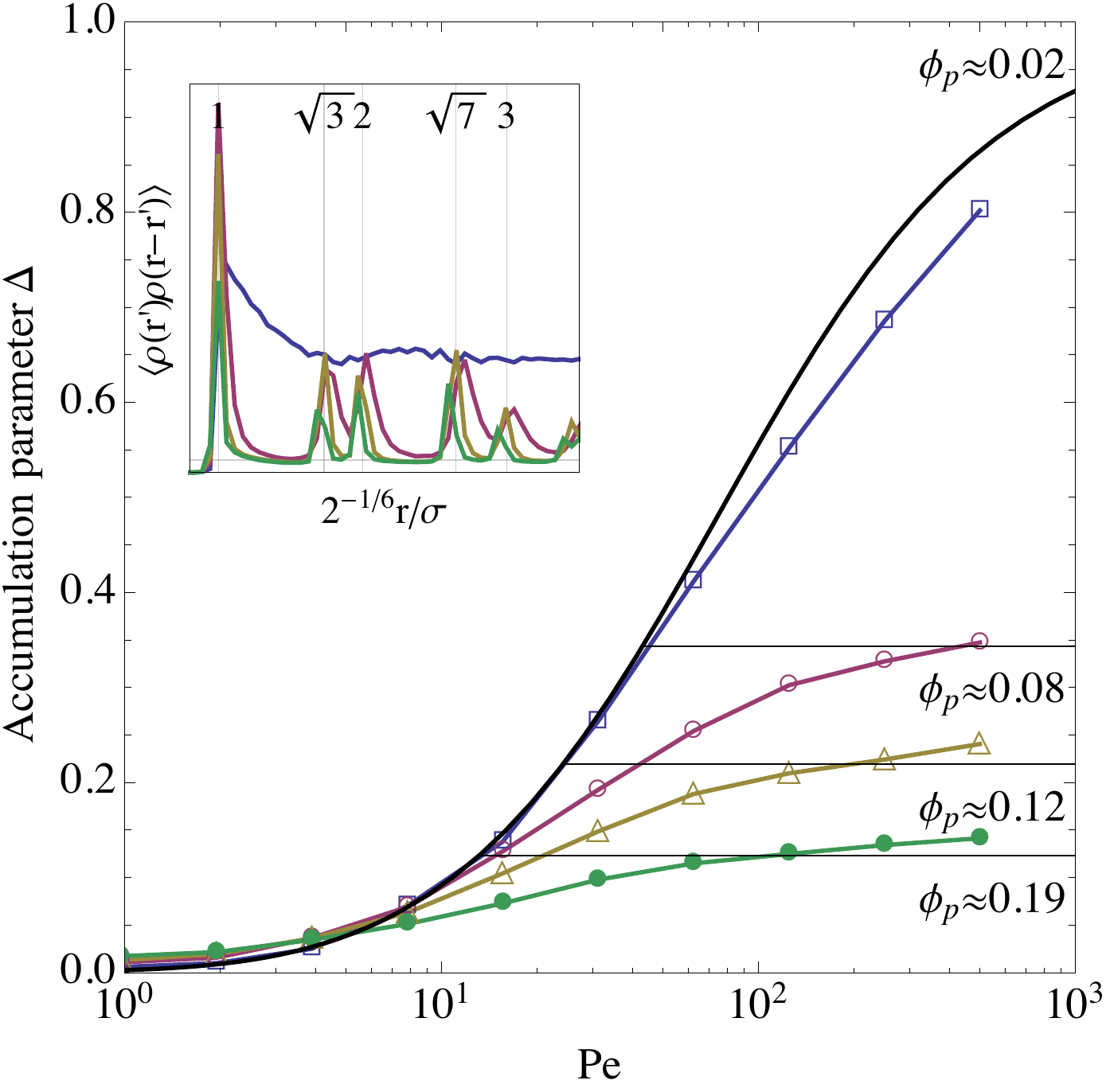}
\caption{\label{fig:Delta_interacting}Accumulation for interacting
  active colloidal particles with volume fractions $\phi_p$ and a
  passive region of relative size $\phi=\pi/100$. The horizontal lines represent
  accumulation parameters for the case $\phi_{p, \mathcal P} = \pi/(2
  \sqrt{3})$ with lattice constant $2^{1/6}\sigma$, which is only an
  upper estimate, as higher densities produce smaller lattice
  constants. Interaction effects are expected for densities exceeding
  $\phi_p^\mathrm{max} \approx 0.028$, where passivated particles align
  in a hexagonal lattice. In the inset the pair correlation function
  for $\Pe = 250$ is shown. The shift of the peaks to reduced
  distances with increasing densities represents a reduction of the
  lattice constant.}
\end{figure}
In Fig.~\ref{fig:Delta_interacting} we observe the reduction of
accumulation for $\phi_p>\phi_p^\mathrm{max}$. Using the hexagonal
closed packing particle density in the passive region (cf.\
Fig.~\ref{fig:sketch}) gives a good estimate for the accumulation
parameter in the high-$\Pe$ limit,
\begin{equation}
\Delta_\mathrm{max}(\phi_p, \phi) = \frac{\phi}{1-\phi}\left ( 1-\frac{\frac{\pi}{2 \sqrt{3}}}{\phi_p}\right ).
\end{equation}
For high particle densities,
smaller lattice constants compared to the truncation length, leading
to an increased accumulation compared to our estimate. 
 
\begin{figure*}[bt]%
\includegraphics[width=\columnwidth]{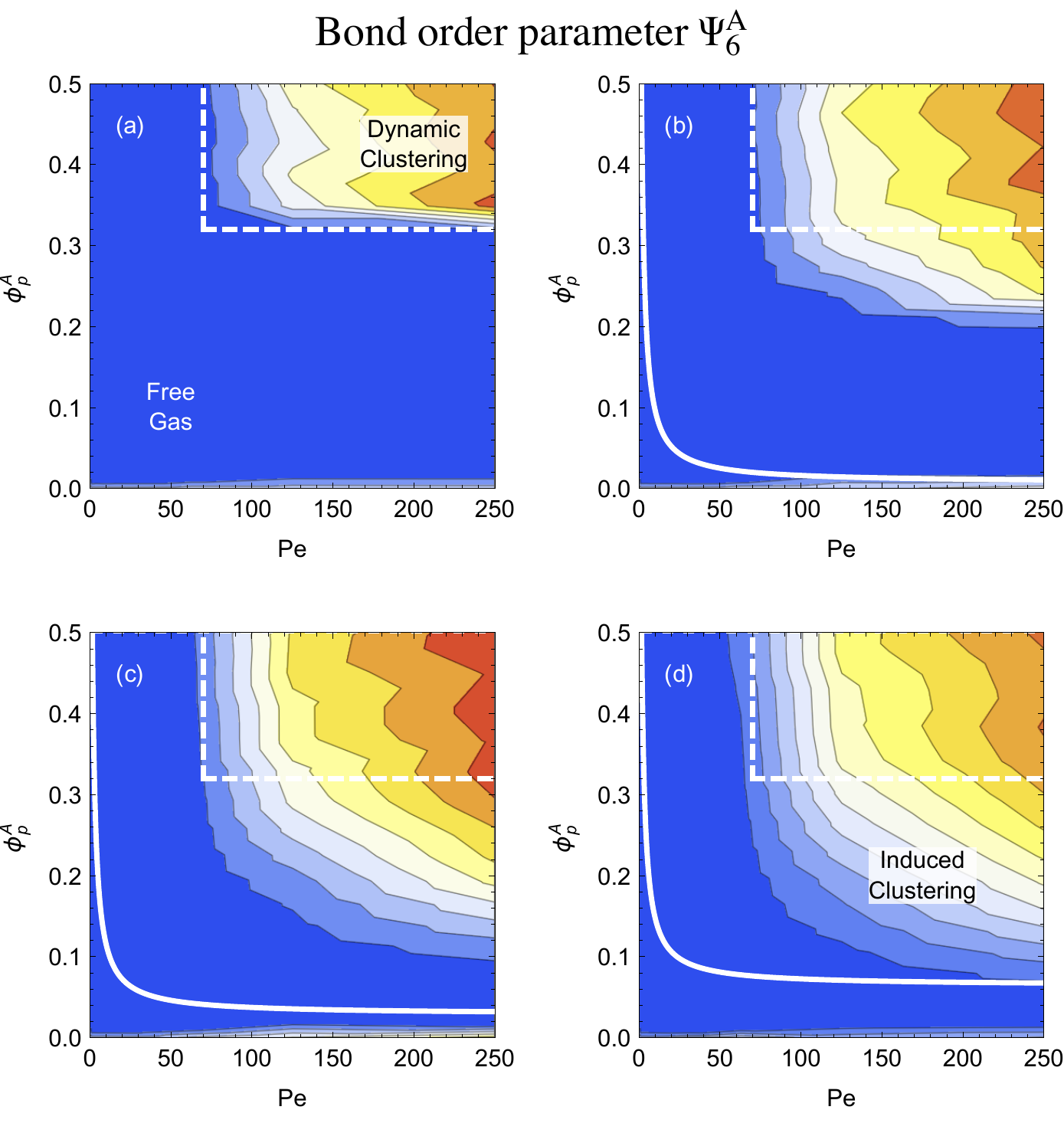}
\includegraphics[width=\columnwidth]{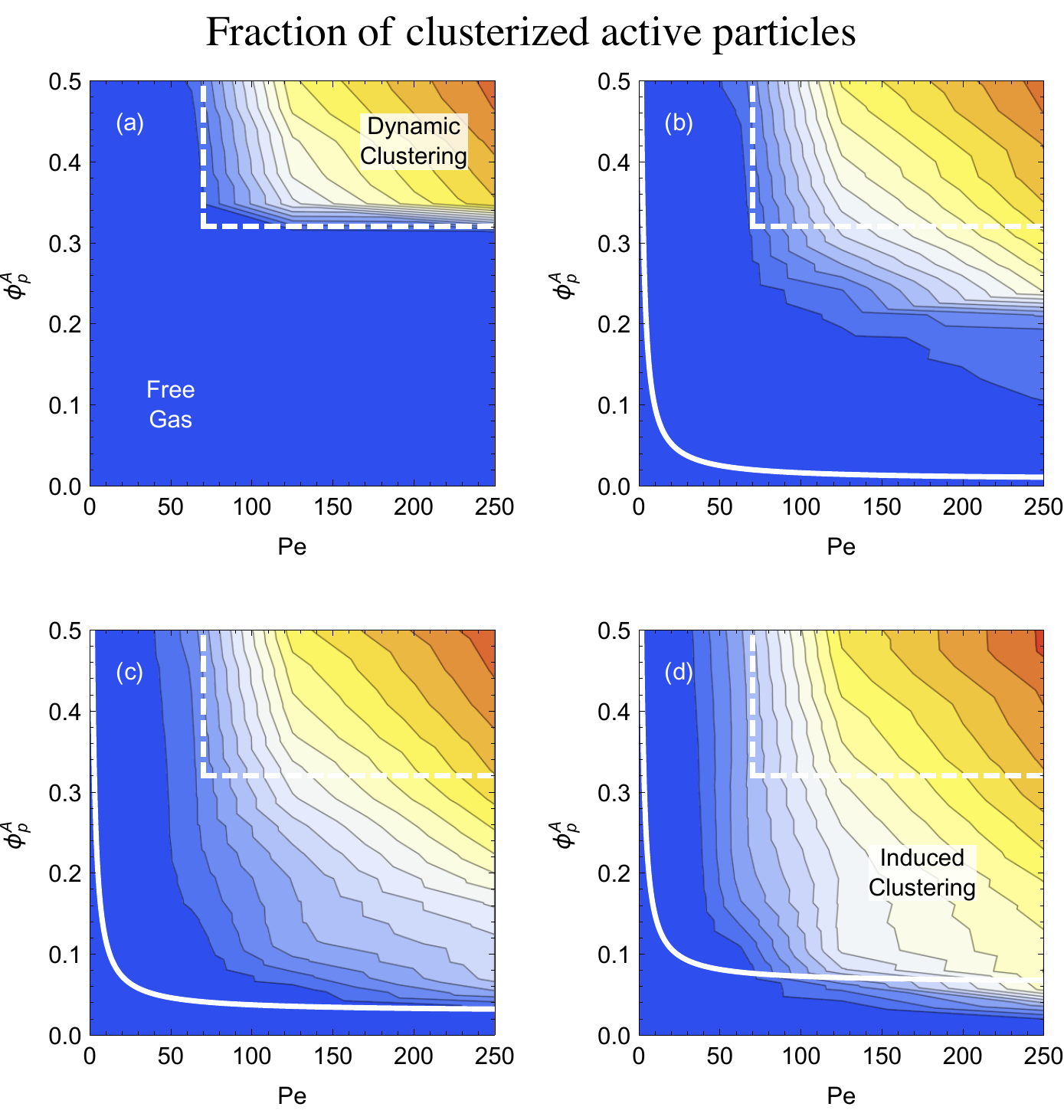}
\includegraphics[width=.75
\columnwidth]{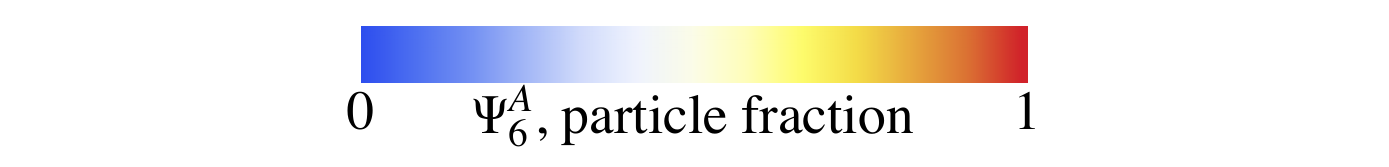}%
\caption{\label{fig:phases}Phase diagram depending on activity and
  density of active particles in the active region, expressed by the \emph{active}
  bond order parameter as well as the clustered fraction of particles for (a) $\phi=0$ (no passive region) and (b-d)
  $\phi = \pi /400, \pi/100, 9 \pi/400$. The white dashed square serves as a
  guide to the eye. The white curve indicates at which filling
  fraction the passive region is filled assuming a
  lattice constant $2^{1/6}\sigma$.}%
\end{figure*}%
\subsection{High particle density: Dynamic clustering}
Above, we focussed on an accumulation in the passive region. But how
is the interplay between this diffusion-stimulated accumulation
discussed so far and spontaneous clustering which finally leads to the
athermal phase separation \eg observed at $\phi_p {\gtrsim} 0.4$ in
Ref.~\cite{Fily2012}? Therefore, we first recapitulate the responsible
mechanism for the latter: Active particles which collide can get stuck
and create a cluster, if they are oriented towards each other. We
define a cluster as a composition of at least three particles, with
distances less then $2^{1/6} \sigma$. A dynamically
created cluster survives until a member of the cluster changes its
orientation through orientational Brownian motion and evades, which
takes some time $\pt$. If within this time new particles have
condensed at the cluster surface, inner particles remain until
the outer ones reorient, leading to the emergent accumulation of one
big cluster. This happens in the phase separated case for active
particles at high activities and high densities, and is indicated in
Fig.~\ref{fig:phases} (a) by an increased bond-order parameter which
reads for a hexagonal lattice \cite{Nelson1979,Schmidle2012}
 \begin{equation}
\Psi_6 = \frac{1}{N} \bigg | \sum_{i} \psi_6(i) \bigg |
   \quad \mathrm{with} \quad
 \psi_6(i)= \frac{1}{6} \sum_{j} e^{i 6 \theta_{ij}},
\label{eq:Psi6}%
  \end{equation} 
with $\theta_{ij}$ being the bond angle between particle $i$ and $j$
and an arbitrary reference axis. If particle $i$ is surrounded by
six nearest neighbors in a hexagonal manner, the local order
  parameter $|\psi_6(i)|$ is 1.  $\Psi_6^\light$ denotes the global
  order parameter for the active region only.  Accordingly in
  Fig.~\ref{fig:sketch}, we observe many passivated particles with
  $|\psi_6(i)|{=}1$.

If a passive region is introduced, the clustering effect in the active
region is influenced by a reduction of the particle density due to
partial capturing of particles in the passive region. We account for
this effect by calculating quantities with respect to the particle
volume fraction in the active region only. Furthermore, we observe in
Fig.~\ref{fig:sketch} that the diffusion-induced cluster grows beyond
the boundary of the passive region, leading to cluster formation also
in the active region at particle densities where the athermal phase
separation in a purely active system does not take place. This effect
is related to the athermal phase separation as well as the
accumulation of active particles \cite{Volpe2011} or bacteria near
walls \cite{Takagi2014}: Active particles impinging at the surface of
the passively induced cluster first have to reorient, which takes some
persistence time $\pt$. As the collision partner of an impinging
particle cannot evade (because it is stuck in the passive region), the
rest time of an impinging particle is longer compared to that in a
dynamically induced cluster.  Thus, the passive region acts as a
nucleation core for dynamic clustering, which can now take place at
reduced densities.

In Figs.~\ref{fig:phases}(b-d) we can observe a reduction of the
minimal clustering density depending on the size of the passive
region. Not shown here is a pinning effect: While in a purely active
system the clusters diffuse  freely, a passive region introduces
a pinning site for the cluster. This makes the introduction of a
passive region, in combination with the dynamic clustering, interesting
for applications which require control of a target region of active
particles.

\section{Conclusion and outlook}
In this work a method to trap propelled colloidal particles is
presented. It is based on the suppression of the propulsion mechanism,
\egg, the introduction of a shading mask in a system of phoretically
propelled particles. An accumulation parameter is defined on the basis
of inhomogeneous diffusion for diluted systems, and confirmed by
computer simulations. A finite size effect is pointed out, which
explains why the defined parameter is only valid as an upper
limit. For the case of increased particle densities, the impact of
steric interactions between the active particles has been adressed,
which leads to a decrease of the accumulation as soon as
crystallization in the passive region occurs. A further increase of
the particle density leads to the regime where dynamic cluster
formation sets in. We observe an enhancement of the effect of athermal
phase separation by the introduction of a passive region. The
combination of trapping by passivation with dynamic clustering makes
the emergent clusters more controllable: Targets can be marked by
introducing an inhomogeneity into the system. The collective behavior
of the active particles then leads to a strong concentration of active
particles in the \emph{target region}, which may fulfill their
task. When the mark is removed again, the active particles can again
disperse in the medium.

An important next step is the experimental verification of the results
presented here. Concerning the case of dense systems,
experiments already exist: In a system of \textit{Escherichia Coli}
bacteria \cite{Demir2012} the system cell has been manipulated such
that bacteria move with a different speed in different regions of the
system, leading to their accumulation in the region with lower
velocity.  In another experiment, the inverse effect has been observed
in a system of SiO$_2$-Ag Janus particles in a H$_2$O$_2$ solution
\cite{Sen2009}: Irradiation with UV-light causes a photolysis reaction
between Ag and H$_2$O$_2$ and finally results in a
self-diffusiophoretic motion of the particles. The fuel-gradient
causes a diminishing particle concentration towards the center of the
UV spot.  This effect even dominates phototaxis of the particles,
which otherwise would generate an accumulation of particles in the
spot. 

Besides experimental comparison several other questions remain. The
perhaps most important one is that of the impact of interactions
between the particles. Hydrodynamic interactions between active
particles have been neglected so far. However, it is known that
hydrodynamic interactions can strongly influence the collective
behavior of active colloids. In \cite{Hennes2014} for example, the
existence of a harmonic trap in a system of active particles leads to
the formation of a pump or a rectifier, as the hydrodynamic interaction
leads to a parallel alignment of particles approximating particles. A
similar effect can be expected for a system of active particles in an
inhomogeneous medium.
\begin{acknowledgments}
  M.M.\ thanks C.~Bechinger and F.~K\"ummel for stimulating
  discussions on dynamical clustering and particles in patterned
  media, which led to the study of higher densities of active
  particles.  The authors thank Dietrich E.\ Wolf for discussions.  A
  grant by the University of Duisburg-Essen for the promotion of young
  scientists is acknowledged.
\end{acknowledgments}

%\bibliography{library}

%

\end{document}